\documentstyle[preprint,aps,epsf]{revtex}
\begin{document}
\title{Nonlinear saturation of electrostatic waves:\\ mobile ions modify
trapping scaling}
\author{John David Crawford and Anandhan Jayaraman}
\address{Department of Physics and Astronomy\\
University of Pittsburgh\\
Pittsburgh, Pennsylvania  15260}
\date{June 24, 1996}
\maketitle
\begin{abstract}
The amplitude equation for an unstable electrostatic wave in a multi-species
Vlasov plasma has been derived. The dynamics of the mode amplitude $\rho(t)$ is
studied using an expansion in $\rho$; in particular, in the limit
$\gamma\rightarrow0^+$, the singularities in the expansion
coefficients are analyzed to predict the asymptotic dependence of the electric
field on the linear growth rate $\gamma$. Generically $|E_k|\sim \gamma^{5/2}$,
as $\gamma\rightarrow0^+$, but in the limit of infinite ion mass or for
instabilities in reflection-symmetric systems due to real eigenvalues the more
familiar trapping scaling $|E_k|\sim \gamma^{2}$ is predicted.
\end{abstract}


The evolution of an unstable electrostatic mode is a fundamental
problem in collisionless plasma theory, and is perhaps the simplest
nonlinear problem requiring a self-consistent treatment of the resonant
interaction between waves and particles. When resonant particles interact
with a large amplitude wave, then much of the behavior can be understood by
analyzing the particle motion as if the wave amplitude were constant;
this approximation linearizes the problem.\cite{on1} Alternatively, if the wave
amplitude is sufficiently small then the initial instability can be predicted
treated by neglecting the effect of the wave on the particles; this leads to
conventional linear Vlasov theory. However, to describe the dynamics of the
unstable mode
which develops from a small initial amplitude into a final nonlinear state
requires an analysis of the self-consistent and nonlinear interaction between
the wave and the resonant particles.

{}From a dynamical systems viewpoint even the simplest examples of
instabilities
in a Vlasov plasma have many unusual features related to the  Hamiltonian
character of the dynamics and the central role played by the neutrally stable
continuous spectrum in the appearance of the unstable modes.\cite{cra1} These
novelties are not present solely in Vlasov theory; entirely analogous features
arise in models of unstable inviscid shear flows, in stability calculations for
certain classes of solitons, and in theories of large systems of coupled
oscillators.\cite{case2}-\cite{sm} In the better understood setting of
dissipative systems, the nonlinear evolution of the mode amplitudes can be
described using an expansion in the amplitude of the unstable modes. Since the
growth rates are very small near onset, nonlinear effects often act to saturate
the instability before the amplitudes grow appreciably; for this reason such
expansions have proved a powerful tool for studying the nonlinear states
emerging from the bifurcation.\cite{jdc2,crossh}

It has long been hoped that similar methods could be applied to the Vlasov
equation despite the absence of dissipation. However, for many years, efforts
to construct such expansions, even for the case of a single unstable mode, have
been plagued by the fact that the nonlinear terms involved divergent integrals
due to resonant denominators. Thus the calculations appeared to breakdown
precisely in the regime where the amplitudes of the unstable waves were
extremely small. In addition, efforts to regularize the the expansion
coefficients inevitably led to theories that predicted scaling behavior for the
saturated amplitudes that contradicted numerical
results.\cite{frieman}-\cite{bald}
More precisely, these theories predicted that the electric field of the
saturated mode would satisfy $E\sim\sqrt{\gamma}$ as $\gamma\rightarrow0^+$,
whereas numerical simulations find the exponential growth of the mode halted at
an amplitude characterized by the ``trapping scaling''
$E\sim\gamma^2$.\cite{dru}-\cite{sim2}

Recently, we have made  progress on this problem for the Vlasov equation; both
in the approach to constructing the expansions and in the way the singular
limit $\gamma\rightarrow0^+$ is treated and
interpreted.\cite{cra4}-\cite{jdc95} An amplitude equation for an unstable mode
in a one-dimensional collisionless plasma was derived for
the dynamics on the two-dimensional unstable manifold of the equilibrium $F_0$.
The essential difference from previous work lies in the choice of unperturbed
state. Earlier theories assumed an equilibrium with a neutrally stable mode and
obtained ill-defined expansion coefficients.\cite{sim1} This can be avoided by
taking the weakly unstable equilibrium as the unperturbed state; a choice that
naturally leads one to work with the unstable manifold. The mode eigenvalue
$\lambda=\gamma-i\omega$ can be complex (beam-plasma) or real (two-stream); in
either case the equations for the amplitude $A(t)=\rho(t)\,e^{-i\theta(t)}$,
\begin{eqnarray}
\dot{\rho}&=&\gamma\rho+a_1\rho^3+a_2\rho^5+{\cal O}({\rho^7})\label{hopf1}\\
\dot{\theta}&=&\omega+{a'_1}\rho^2+{a'_2}\rho^4+{\cal O}({\rho^6})\label{hopf2}
\end{eqnarray}
lead to a one-dimensional problem for $\rho$ because the
spatial homogeneity of the equilibrium decouples the phase dynamics.
As $\gamma\rightarrow0^+$, the expansion coefficients diverge,
\begin{equation}
a_j,\, {a'_j} \sim\frac{1}{\gamma^{4j-1}},\label{divold}
\end{equation}
but these divergences can be removed to {\em all} orders in the expansion by
rescaling the mode
amplitude: $\rho(t)\equiv\gamma^2\,r(\gamma t)$. In this way one obtains
asymptotic equations for $r(\tau)$ that are well behaved as
$\gamma\rightarrow0^+$, and moreover through Poisson's equation
this rescaling implies that the electric field exhibits the trapping scaling.
These initial results were obtained for a plasma of mobile electrons with
infinitely massive ions providing a fixed neutralizing background; consequently
they contain no information regarding unstable ion acoustic modes, for example.

In order to study the effects of
ion dynamics, we have generalized the analysis to treat an unstable
electrostatic mode in a {\em multi-species} one-dimensional plasma. This
changes the problem in a qualitative way: now as $\gamma\rightarrow0^+$, the
single particle dynamics and collective motions of the ions occur on a fast
time scale relative to $1/\gamma$. We calculate explicit expressions for the
leading nonlinear
coefficients $a_1$ and $a'_1$; in addition we have determined the dominant
singularities in the amplitude expansion to all orders.
The results show a  qualitatively different
singularity structure from the limiting model (\ref{divold}) with fixed ions,
and provide new predictions for the scaling of nonlinearly saturated modes.

The theory is described for the simplest example of a neutral plasma with two
species $(s=e,i)$ which we refer to as ``electrons'' and ``ions'' although the
results apply equally to collisionless electron-positron plasmas. Let
$n_s=N_s/L$ denote the average species density in a one-dimensional plasma of
length $L$, and $eq_s$ denotes the charge per particle of species $s$. In
convenient dimensionless variables, the Vlasov-Poisson system becomes
\begin{equation}
\partial_t F^{(s)}+v\partial_x F^{(s)}+\kappa^{(s)}E\,\partial_v
F^{(s)}=0\hspace{0.25in}
\partial_xE=\sum_s\int^\infty_{-\infty}dv\,F^{(s)}\label{vlasov}
\end{equation}
where $\kappa^{(s)}\equiv(q_sm_e/m_s)$.
We assume periodic boundary conditions and adopt the normalization
\begin{equation}
\int^{L/2}_{-L/2}dx\,\int^\infty_{-\infty}dv\,F^{(s)}(x,v,t)=
\frac{q_sn_s}{n_e}L;
\label{norm}
\end{equation}
note that $F^{(s)}$ is negative for electrons and positive for ions.
Given a spatially homogeneous equilibrium $F^{(s)}_0(v)$, this system
determines an evolution equation for $f^{(s)}(x,v,t)\equiv
F^{(s)}(x,v,t)-F^{(s)}_0(v)$
\begin{equation}
\partial_t f^{(s)}={\cal{L}} f^{(s)}+{\cal{N}}(f^{(s)})\label{pertvlasov}
\end{equation}
where
${\cal{L}} f^{(s)}=-v\partial_x f^{(s)}-\kappa^{(s)}E\,\partial_v F^{(s)}_0$
and ${\cal{N}}(f^{(s)})=-\kappa^{(s)}E\,\partial_v f^{(s)}$.

An unstable mode exists if the
dielectric function
\begin{equation}
\epsilon_{{k}}(z)\equiv 1- \frac{1}{k^2}\int^\infty_{-\infty}\,dv\,
\frac{\sum_s\kappa^{(s)}\partial_vF^{(s)}_0(v)}{v-z}
\hspace{0.5in}(\mbox{\rm Im}\;z>0)
\label{diefcn}
\end{equation}
has a root $z_0=v_p + {i\gamma}/{k}$ in the upper half plane $(\gamma>0)$.
The root determines an eigenvalue $\lambda=-ikz_0$ for ${\cal{L}}$ with a
two-component eigenvector
\begin{equation}
\Psi=e^{ikx}
\left(\begin{array}{c}\psi^{(e)}(v)\\ \psi^{(i)}(v)\end{array}\right).
\label{efcn}
\end{equation}
We assume that there is a single such mode and that it corresponds to a simple
root of $\epsilon_{{k}}(z)$, i.e. $\epsilon_{{k}}'(z_0)\neq0$.
The eigenvalue $\lambda$ can be real or complex depending on
the equilibrium; the ion-acoustic instability corresponds to a complex
eigenvalue.

The amplitude of the unstable mode is the coefficient of $\Psi$ in
the expansion of $f$
\begin{equation}
f(x,v,t)=\left[A(t)\Psi(x,v) + cc\right] + S(x,v,t);\label{linmodes}
\end{equation}
here $f$ denotes the two-component field $f\equiv(f^{(e)},f^{(i)})$.
This decomposition allows the dynamics of the mode amplitude $A(t)$ and the
remaining modes $S(x,v,t)$ to be separated
\begin{eqnarray}
\dot{A}&=&\lambda\,A+(\tilde{\Psi},{\cal{N}}(f))\label{Adot}\\
\partial_t S&=&{\cal{L}} S+{\cal{N}}(f)-\left[(\tilde{\Psi},{\cal{N}}(f))\,\Psi
+ cc\right]
\label{Sdot}
\end{eqnarray}
using an inner product defined  for two-component fields $B=(B^{(e)},B^{(i)})$
and $D=(D^{(e)},D^{(i)})$ by
$(B,D)\equiv\int\,dx\int\,dv\,[ B^{(e)}(x,v)^\ast D^{(e)}(x,v) +
B^{(i)}(x,v)^\ast D^{(i)}(x,v)]$, and the adjoint eigenfunction $\tilde{\Psi}$
for $\lambda^\ast$.
In (\ref{Adot})-(\ref{Sdot}) the linear terms are decoupled, but nonlinear
couplings between $\dot{A}$ and $\partial_t S$ remain.

The amplitude equation for ${A}$ follows when
we express the time dependence of $S$ in terms of ${A}$:
\begin{equation}
S(x,v,t)=H(x,v,A(t),A^\ast(t)).\label{graph}
\end{equation}
As we have discussed elsewhere, this step can be visualized as a restriction
of the initial condition to the two-dimensional unstable manifold of the
equilibrium.\cite{jdc94,jdc95} Consistency between the time dependence of $S$
in (\ref{graph}) and the evolution of $S$ described by
(\ref{Adot})-(\ref{Sdot}) requires that $H(x,v,A(t),A^\ast(t))$ satisfy
\begin{equation}
\left[\dot{A}\,\partial_{A}H+\dot{A}^\ast\,\partial_{A^\ast}H\right]
{\rule[-3.0mm]{0.25mm}{8.0mm}}_{f=f^u}
={\cal{L}} H+{\cal{N}}(f^u)-\left[(\tilde{\Psi},{\cal{N}}(f^u))\,\Psi +
(\tilde{\Psi},{\cal{N}}(f^u))^\ast\,\Psi^\ast\right]\label{Heqn}
\end{equation}
where $f^u(x,v)=\left[A\Psi(x,v) + cc\right] + H(x,v,A,A^\ast)$. For solutions
of this form,  the dynamics of $A(t)$ (\ref{Adot}) yields an
autonomous equation for $A$
\begin{equation}
\dot{A}=\lambda\,A+(\tilde{\Psi},{\cal{N}}(f^u))\label{unAdot}
\end{equation}
provided $H$ can be determined from (\ref{Heqn}).
The homogeneity of the equilibrium $F_0$ forces this amplitude equation to have
a simple form: $\dot{A}=A\,p(|A|^2)$ where $p(|A|^2)$ must still be determined.
In polar variables, $A=\rho e^{-i\theta}$, the system (\ref{unAdot}) separates
\begin{equation}
\dot{\rho}=\rho\,\mbox{\rm Re}\,[p(\rho^2)]\hspace{0.75in}
\rho\dot{\theta}=-\mbox{\rm Im}\,[p(\rho^2)]\label{amphase}
\end{equation}
yielding a one-dimensional flow for $\rho(t)$; the essential problem is to
study $p(\rho^2)$.

Our conclusions regarding the evolution of the wave are based on an analysis
of the amplitude expansion for $p$,
\begin{equation}
p(\rho^2)=\sum^\infty_{j=0}\,p_j\,\rho^{2j},\label{series}
\end{equation}
and similar expansions for $H(x,v,A,A^\ast)$. By substituting
$\dot{A}=A\,p(\rho^2)$ into (\ref{unAdot}) we obtain one set of relations
between the coefficients of $p$ and $H$:
\begin{equation}
A\sum^\infty_{j=0}\,p_j\,\rho^{2j}=\lambda\,A+(\tilde{\Psi},{\cal{N}}(f^u));
\label{eq:pcoeff}
\end{equation}
the defining equation (\ref{Heqn}) for $H$ provides a second set of relations.
The expansion coefficients are calculated by solving (\ref{Heqn}) and
(\ref{eq:pcoeff}) recursively; full details of this calculation
and our analysis of the
resulting recursion relations will be published elsewhere.\cite{cj96b}

The first coefficient $p_0$ is simply the linear eigenvalue $\lambda$;
information
about the nonlinear evolution is provided by the higher order coefficients,
in particular the first nonlinear term is given by
\begin{equation}
p_1=\frac{1}{\gamma^4}
\left[\frac{\kappa^{(i)}(1-{\kappa^{(i)}}^2){\rm
Im}(\alpha(z_0))}{4k\epsilon_{{k}}'(z_0)}
-\gamma\left(\frac{1}{4}+
\frac{\kappa^{(i)}(1-{\kappa^{(i)}}^2)\alpha'(z_0)}
{4k^2\epsilon_{{k}}'(z_0)}\right)+
{\cal O}(\gamma^2)\right]\label{eq:p1}
\end{equation}
where $\alpha'(z)$ is the derivative of
$\alpha(z)\equiv\int dv\partial_v F_0^{(i)}/(v-z)$. If the first term in
(\ref{eq:p1}) is non-zero, i.e.
\begin{equation}
\kappa^{(i)}(1-{\kappa^{(i)}}^2){\rm Im}(\alpha(z_0))\neq0,\label{eq:newterm}
\end{equation}
then in the limit of weak instability $\gamma\rightarrow0^+$, there is
a $\gamma^{-4}$ singularity in $p_1$. When (\ref{eq:newterm}) holds, the
trapping scaling
introduced previously for $\rho$ no longer removes the singularity in $p_1$,
and a new scaling $\rho(t)=\gamma^{5/2}\,r(\gamma t)$ is required to obtain a
nonsingular leading order term.

Beyond this leading order calculation,
we have proved that the coefficients in (\ref{series}) have the asymptotic
behavior $p_j\sim\gamma^{-(5j-1)}$ to all orders $j\geq1$. This general result
implies that when the equations (\ref{amphase}) are rewritten in the rescaled
amplitude $r$ the singularities will be cancelled
to {\em all} orders in  $r$.\cite{cj96b} Thus given the generic condition
(\ref{eq:newterm}) the theory predicts the asymptotic scaling for the mode
electric field is $|E_k|\sim \gamma^{5/2}$ as $\gamma\rightarrow0^+$.

Our prediction must be re-examined if the genericity condition
(\ref{eq:newterm}) fails and the leading term in $p_1$ vanishes. The three
factors in (\ref{eq:newterm}) indicate three exceptional circumstances when
this can happen: $\kappa^{(i)}=0$, ${\rm Im}(\alpha(z_0))=0$ or
${\kappa^{(i)}}^2=1$. The first circumstance corresponds to the limit of fixed
ions, $m_e/m_i\rightarrow0$, and the cubic term (\ref{eq:p1}) reduces to our
previous result (\ref{divold}). With fixed ions, the terms of the amplitude
expansion are known to be
nonsingular to all orders once the amplitude has been scaled with $\gamma$ to
balance the divergence in $p_1$, i.e. $\rho(t)=\gamma^2\,r(\gamma
t)$.\cite{jdc95}

The second circumstances arises naturally when the equilibria under
consideration are
reflection symmetric $F^{(s)}_0(-v)=F^{(s)}_0(v)$.
With such a reflection symmetry, one can find pure imaginary roots
$z_0^\ast=-z_0$ (hence real eigenvalues) and then $\alpha(z_0)$ is forced to
be real; an example is provided by a reflection-symmetric two-stream
instability. In this case, the trapping scaling for $\rho$ cancels the
singularity in $p_1$ and moreover can be proved to yield
an amplitude expansion which is
nonsingular to all orders.\cite{cj96b} As in the fixed ion case, this result
predicts the familiar trapping scaling for the mode electric field
$|E_k(t)|\sim \gamma^2$.

The third exceptional circumstance
${\kappa^{(i)}}^2=1$ requires $(q_i m_e/m_i)^2=1$ which
corresponds to an electron-positron plasma ($q_i=1$, $m_i=m_e$). In this case
the singularity structure of the expansion is more complicated. The
cubic coefficient has a $\gamma^{-3}$ singularity which suggests trapping
scaling for $\rho$, but the divergence of fifth order coefficient turns out to
be $p_2\sim\gamma^{-8}$ which is {\em not} removed by trapping
scaling.\cite{cj96b}
This fifth
order singularity is cancelled if we set $\rho(t)=\gamma^{9/4}\,r(\gamma t)$.
However, inspection of the higher order terms in the expansion shows the
divergence structure of $p_j\sim\gamma^{-(5j-3)}$ and cancelling these
singularities to all orders again requires the generic scaling
$\rho(t)=\gamma^{5/2}\,r(\gamma t)$ even though to any finite order $p_j$ a
smaller exponent would suffice.\cite{cj96b}

With the notable exception of this third example,
our conclusions are easily summarized: the scaling required to obtain a
nonsingular expansion is correctly predicted by the divergence found in the
cubic coefficient $p_1$. Generically, this singularity dictates a scaling by
$\gamma^{5/2}$, but this is replaced by $\gamma^{2}$ in the limit of infinite
ion inertia or for instabilities in reflection-symmetric systems due to real
eigenvalues. In the
generic case we can estimate the range of growth rates where the new
$\gamma^{5/2}$ scaling is visible by determining the range in $\gamma$
where the $\gamma^{-4}$ divergence dominates $p_1$.  In general, this will
depend on the specific parameters of the equilibrium, and we give two
illustrative examples: an unstable plasma wave driven by an electron beam and
an ion-acoustic instability. For simplicity, we have evaluated the dispersion
relation and the cubic coefficient using Lorentzian distributions for the
two species,
\begin{equation}
F_0^{(e)}(v)=\frac{-{\alpha_e}\,n_p/{\pi}}{(v-u_e)^2+\alpha_e^2}-
\frac{{\alpha_b\,n_b}/{\pi}}{(v-u_b)^2+\alpha_b^2}
\hspace{0.35in}
F_0^{(i)}(v)=\frac{\kappa^{(i)}{\alpha_i}/{\pi}}{v^2+\alpha_i^2},
\label{eq:family}
\end{equation}
with $n_p+n_b=n_i=1$.

\begin{figure}
\vspace{0.0in}
\begin{center}
\leavevmode
\hbox{%
\epsfxsize=4.0in
\epsffile{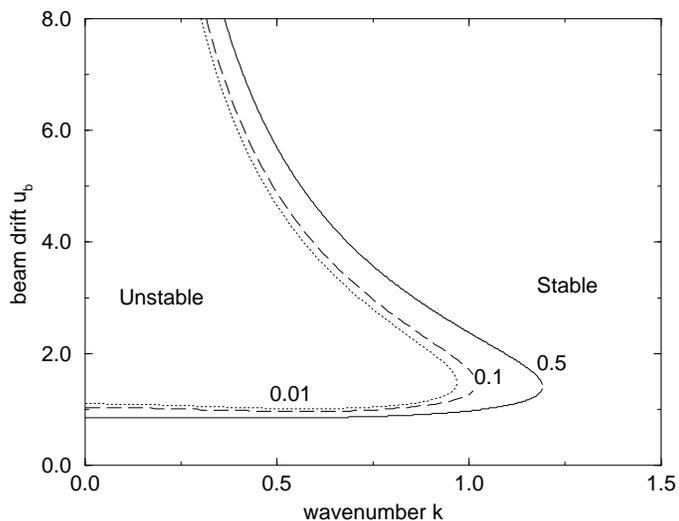}}
\end{center}
\caption{{\bf Beam-plasma stability boundary for $\alpha_e=1.0$, $\alpha_i=1.0$
and $\alpha_b=0.1$ at four mass ratios $m_e/m_i=0.001, 0.01, 0.1$ and $0.5$.
The boundaries for
$m_e/m_i=0.001$ and $0.01$ coincide.}}
\label{fig:fig1}
\end{figure}

The plasma wave instability is calculated for an equal density cold beam with
four mass ratios; the linear stability boundaries are shown in Fig. 1. The
densities in this example are $2n_p=2n_b=n_i=1$, and $u_e=0$. The variation of
$\gamma^3{\mbox{\rm Re}}p_1$ as $\gamma\rightarrow0^+$ is shown in Fig. 2. For
each mass
ratio in Fig. \ref{fig:fig1}, we fix $k$ and vary $u_b$; the chosen values are
$k=0.75$ for $m_e/m_i=0.5$ and $k=0.5$ for $m_e/m_i=0.1, 0.01,$ and $0.001$. As
$m_e/m_i$ decreases, the fixed ion
result $\gamma^3{\mbox{\rm Re}}p_1\rightarrow-1/4$ holds down to smaller and
smaller growth rates. Interestingly, the effect of the mobile ions is to shift
the asymptotic sign of ${\mbox{\rm Re}}p_1$ from negative to positive. Since
the higher order terms in the amplitude expansion cannot be neglected, this
does
not automatically imply the onset of subcritical bifurcation but may
nevertheless be significant.

The ion-acoustic instability is shown in Figs. 3 also  for
four mass ratios. In this case $n_b=0$ and there is only a drifting
electron population with $n_p=n_i=1$. In Fig. 4, for each mass
ratio, we fix $k$ and vary $u_e$; the chosen values are $k=1.5, 1.2, 0.75$,
and $0.4$
corresponding to  $m_e/m_i=0.5, 0.1, 0.01,$ and $0.001$, respectively.
Here the effect of decreasing $m_e/m_i$
is to increase the range of the generic scaling as measured in units of the ion
plasma frequency $\omega_i^2=4\pi e^2n_i/m_i$.

The qualitative result of the scaling $|E_k|\sim \gamma^{5/2}$ is to reduce the
electric field of the nonlinearly saturated wave compared to a wave
characterized by trapping scaling. More work is needed to understand how the
mobile ions manage to cut off the growth of the unstable mode at such small
amplitudes.

\acknowledgments
This work supported by NSF grant PHY-9423583.

%This figure produced as a "portrait" not a "landscape": note epsf gives
%same orientation as ghostview and differs by 90 degrees from xvgr.
\begin{figure}
\vspace{0.0in}
\begin{center}
\leavevmode
\hbox{%
\epsfxsize=4.0in
\epsffile{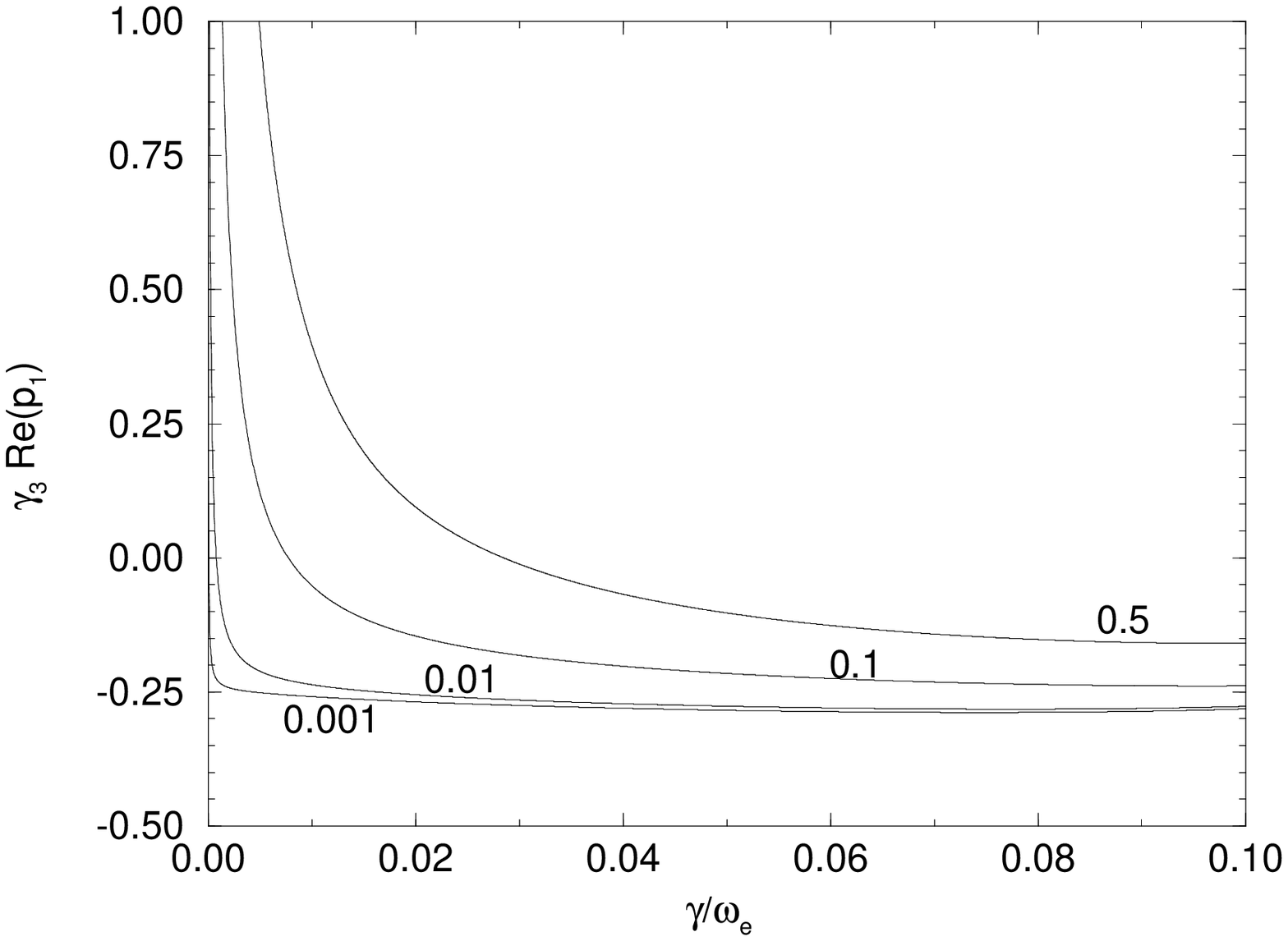}}
\end{center}
\caption{{\bf Asymptotic behavior of $\gamma^3{\mbox{\rm Re}}p_1$ for the
beam-plasma instability of Fig. \ref{fig:fig1}. The growth rate
$\gamma$ is measured in units of the electron plasma frequency.
The divergence as $\gamma\rightarrow0^+$ indicates the regime predicted to show
the generic scaling $|E_k|\sim \gamma^{5/2}$.}}
\label{fig:fig2}
\end{figure}

\begin{figure}
\vspace{0.0in}
\begin{center}
\leavevmode
\hbox{%
\epsfxsize=4.0in
\epsffile{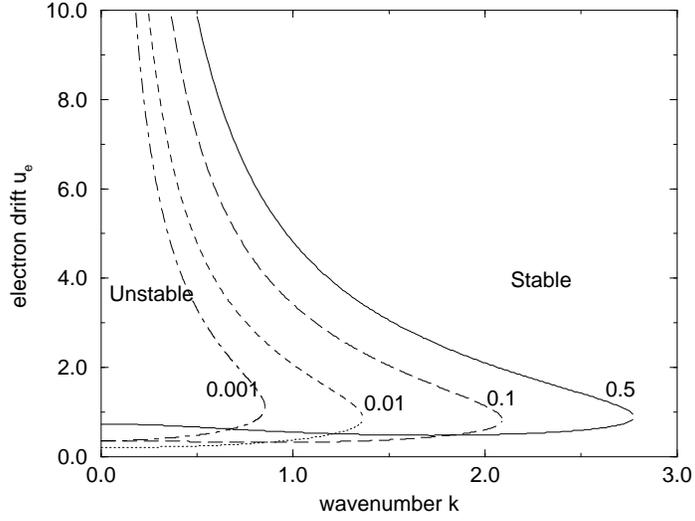}}
\end{center}
\caption{{\bf Ion-acoustic stability boundary for $\alpha_e=1.0$,
$\alpha_i=0.01$ at four mass ratios $m_e/m_i=0.001, 0.01, 0.1$ and $0.5$.}}
\label{fig:fig3}
\end{figure}

\begin{figure}
\vspace{0.0in}
\begin{center}
\leavevmode
\hbox{%
\epsfxsize=4.0in
\epsffile{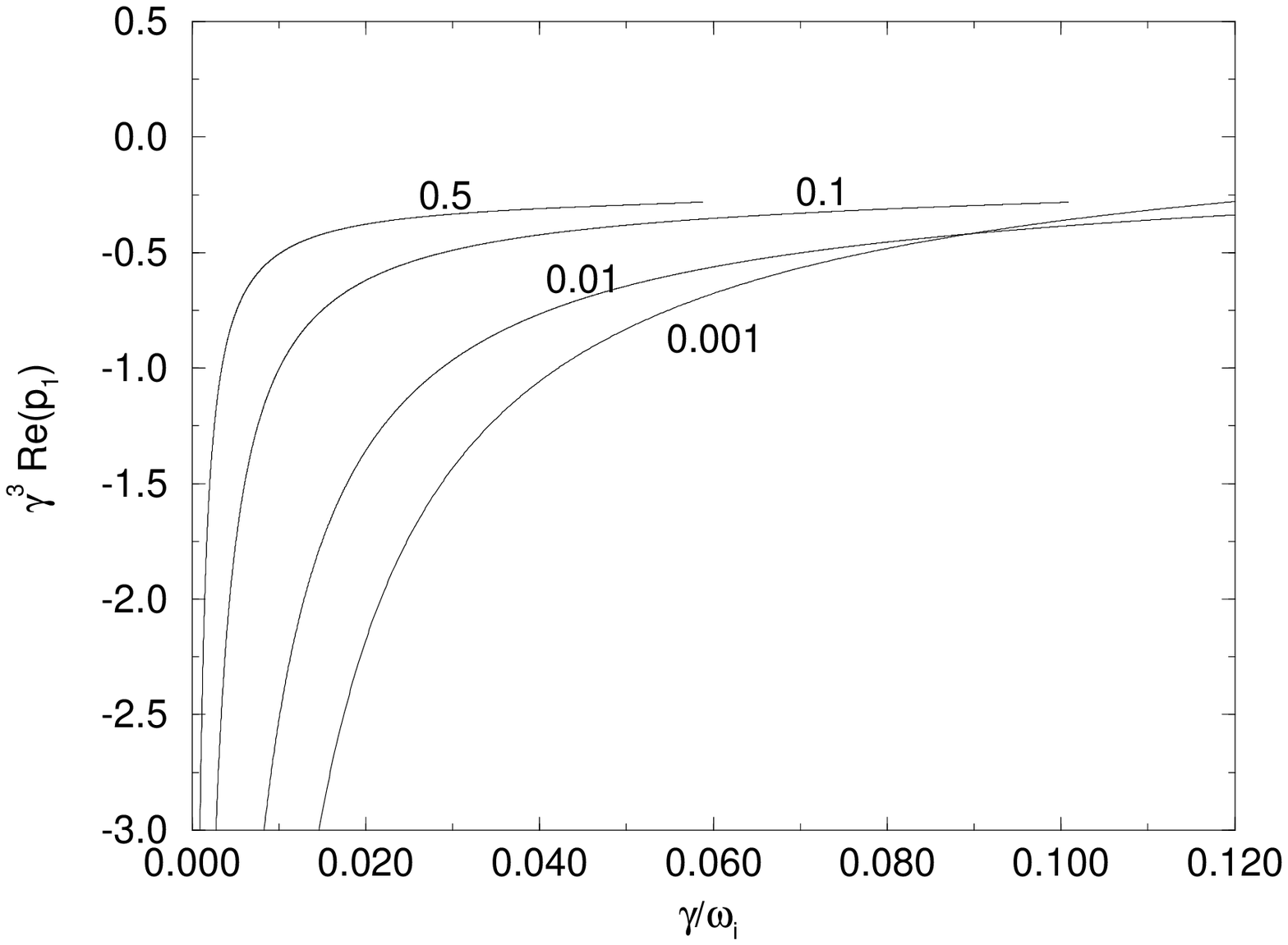}}
\end{center}
\caption{{\bf Asymptotic behavior of $\gamma^3{\mbox{\rm Re}}(p_1)$ for the
ion-acoustic instability of Fig. \ref{fig:fig3}. The growth rate
$\gamma$ is measuredin units of the ion plasma frequency.  The divergence as
$\gamma\rightarrow0^+$, indicates the regime predicted to show the generic
scaling $|E_k|\sim \gamma^{5/2}$.}}
\label{fig:fig4}
\end{figure}

\end{document}